\documentclass[12pt]{iopart}

\usepackage{graphicx} \usepackage{iopams}

\newcommand{\vect}[1]{{\mathbf #1}}
\newcommand{\Frac}[2]{\displaystyle\frac{#1}{#2}}

\begin{document}

\title[Critical States in Disordered Superconducting Films]{Critical
  States in Disordered Superconducting Films}

\author{A. Lamacraft$^{1}$, F. M. Marchetti$^{2}$, J. S. Meyer$^{3}$,
  R. S. Moir$^{2}$ and B. D. Simons$^{2}$}

\address{ $^{1}$Department of Physics, Princeton University,
  Princeton, NJ 08545, USA\\
  $^{2}$Cavendish Laboratory, Madingley Road, Cambridge CB3\ 
  0HE, UK\\
  $^{3}$Department of Physics, University of Minnesota,
  Minneapolis, MN 55455, USA\\
  }

\begin{abstract}
  When subject to a pair-breaking perturbation, the pairing
  susceptibility of a disordered superconductor exhibits substantial
  long-ranged mesoscopic fluctuations. Focusing on a thin film subject
  to a parallel magnetic field, it is proposed that the quantum phase
  transition to the bulk superconducting condensate may be preempted
  by the formation of a glass-like phase with multi-fractal
  correlations of a complex order parameter. Although not universal,
  we argue that such behavior may be a common feature of quantum
  critical phenomena in disordered environments.
\end{abstract}














Several years ago, in an inspiring sequence of papers, it was proposed
by Spivak and Zhou~\cite{spivak_zhou,zhou_spivak} that mesoscopic
fluctuations can significantly influence the nature of the transition
to superconductivity in the vicinity of the upper critical field,
$H_{c2}$. Considering a magnetic field applied perpendicular to a
film, it was argued that the transition to the mixed phase is mediated
by the formation of a superconducting `glass-like' phase realized
through the random Josephson coupling of droplets or domains nucleated
at fields in excess of $H_{c2}$~\cite{spivak_zhou}. Subsequently,
focusing this time on the properties of a superconducting thin-film
subject to a parallel field, qualitatively similar conclusions were
drawn~\cite{zhou_spivak}. Recently, Galitski and
Larkin~\cite{galitski_larkin} extended these ideas exploring the
interplay of the proximity effect and quantum phase fluctuations on
the integrity of the superconducting glass in the perpendicular field
geometry.  Here, focusing on a weakly disordered thin-film subject to
a parallel magnetic field $H$, we will offer a different perspective
on the character of the superconducting phase close to the quantum
critical point. In this geometry, taking the film thickness $d$ to be
smaller than the penetration depth, the field lines enter the sample
and effect a mechanism of pair-breaking leading to the gradual
suppression of the superconducting order parameter. However, in
contrast to Ref.~\cite{zhou_spivak}, we will find it convenient to
limit considerations to the range $d\gg g_L\lambda_F \xi_0/\ell$,
where $g_L$ denotes the Land\'e $g$-factor, $\lambda_F$ is the Fermi
wavelength, $\ell$ the elastic mean free path, and $\xi_0 =
\sqrt{D/\Delta_0}$ (with the classical diffusion constant $D= v_F
\ell/2$) represents the diffusive superconducting coherence length of
the unperturbed system. In this limit, where the impact of Zeeman
splitting can be safely neglected, the transition to the (gapless)
superconducting phase is second order and described by a mean-field
theory of Abrikosov-Gor'kov (AG) type~\cite{abrikosov_gorkov}.

Starting with a microscopic BCS Hamiltonian for the disordered,
symmetry-broken system, one can present the quantum partition function
for an \emph{individual realization} of the impurity potential as a
functional field integral $\mathcal{Z} = \int D[\Delta,\bar{\Delta}]\;
e^{-S_{\Delta}}$, where, to quadratic order in $\Delta$,
\begin{equation}
  S_{\Delta}= \sum_{\omega_m} \int d\vect{r}\, d\vect{r}'\;
  \bar{\Delta}_{\omega_m}(\vect{r})\left[\Frac{1}{\lambda_{\rm BCS}}
  \delta(\vect{r} - \vect{r}') - \Pi_{\omega_m}(\vect{r} , \vect{r}')
  \right] \Delta_{\omega_m}(\vect{r}') \; .
\label{eq:ginzb}
\end{equation}
Here, $\lambda_{\rm BCS}$ denotes the coupling constant of the BCS
interaction while $\Pi_{\omega_m}(\vect{r},\vect{r}') = T
\sum_{\epsilon_n} G_{\epsilon_n} (\vect{r} , \vect{r}')
G_{\omega_m-\epsilon_n} (\vect{r} , \vect{r}')$, where
$G_{\epsilon_n}(\vect{r} , \vect{r}') = \langle \vect{r} |(i\epsilon_n
- \hat{H})^{-1} | \vect{r}'\rangle$ is the sample specific Matsubara
Green function, represents matrix elements of the (generically
complex) Hermitian pairing susceptibility. Since we are primarily
interested in the character of the transition, and not properties deep
in the superconducting phase, the influence of the non-linear terms in
the Hamiltonian can be neglected.

Leaving aside the influence of dynamical fluctuations of the order
parameter, in the saddle-point approximation, the transition to the
superconducting phase is signaled by the appearance of solutions of
the linear equation
\begin{equation}
  \int d\vect{r}'\, \Pi_0(\vect{r} , \vect{r}') \chi(\vect{r}') =
  \nu\gamma\, \chi(\vect{r})\; ,
\label{eq:bcsco}
\end{equation}
with eigenvalues $\gamma\ge (\lambda_{\rm BCS}\, \nu)^{-1}$, defining
$\nu$ as the density of states (DoS) per spin of the normal phase
(assumed constant). In the leading approximation, the influence of
disorder on the symmetry-broken system can be explored by replacing
the pairing susceptibility by its impurity average, $\langle
\hat{\Pi}\rangle_V$. In the specified thin-film geometry, the
`paramagnetic term' due to the parallel magnetic field is negligible,
and $\langle \hat{\Pi}\rangle_V$ takes the form
\begin{equation}
  \langle \Pi_0(\vect{r},\vect{r}') \rangle_V = 4 \pi \nu T
  \sum_{\epsilon_n} \int \Frac{d \vect{q}}{(2\pi)^2}
  \frac{e^{i\vect{q} \cdot (\vect{r} - \vect{r}')}}{D \vect{q}^2 +
  2|\epsilon_n| + 2/\tau_{H}} \; ,
\end{equation}
where $1/\tau_H = De^2(Hd)^2/6$ denotes the pair-breaking rate
associated with the orbital motion of the particle in the external
field~\cite{tinkham}.  In this approximation, when substituted
into~(\ref{eq:bcsco}), one finds that the system becomes unstable
against the formation of a homogeneous condensate when $1/\tau_H$
fulfills the AG condition~\cite{abrikosov_gorkov}, $\ln (T_c/T_c^0) =
\psi (1/2) - \psi (1/2 + 1/2\pi \tau_H T_c)$, where $T_c^0$ is the
transition temperature of the unperturbed system, and
$\psi(z)=\Gamma'(z)/\Gamma(z)$ denotes the digamma function. In
particular, when the pair-breaking rate equals to the order parameter
of the unperturbed system, $2/\bar{\tau}_H^c=\Delta_0 \equiv 2\times
0.88\, T_c^0$, the bulk superconducting phase is destroyed altogether.

However, as emphasized by Spivak and Zhou~\cite{zhou_spivak},
mesoscopic fluctuations of the pairing susceptibility can
significantly influence the character of the transition. In
particular, at the transition, the condensate wave function may
acquire a texture in both \emph{phase} as well as amplitude: Focusing
on the static component, the complex pairing susceptibility exhibits
spatial mesoscopic fluctuations $\delta \Pi_0 (\vect{r} ,\vect{r}') =
\Pi_0(\vect{r} , \vect{r}')-\langle \Pi_0(\vect{r} ,
\vect{r}')\rangle_V$. For a given configuration of the quenched
impurity potential, a system may lower its energy by forming a
condensate where the corresponding wave function is random and
moreover, because of the presence of the magnetic field, complex. In
this case, an arrangement of supercurrent loops can optimally screen
the applied field. In particular, the condensate can exploit the
quenched spatial fluctuations of the pairing susceptibility to
initiate a transition to a superconducting phase ahead of that
predicted by the mean-field estimate above. However, in the bulk
system, such an arrangement of supercurrent loops can be tolerated
only in the near vicinity of the transition. Away from the quantum
critical point, the energy cost in suppressing the amplitude of the
order parameter in the superconducting phase becomes prohibitively
high and a condensate of uniform phase must develop.

To explore the influence of mesoscopic fluctuations on the nature of
the transition one can proceed along complimentary paths: Firstly, one
can develop a scheme perturbative in fluctuations $\delta\Pi$,
treating typical field configurations of the order parameter
$\Delta_0(\vect{r})$ associated with the \emph{impurity averaged
susceptibility} as a variational \emph{Ansatz}. When the scattering
rate $1/\tau_H$ is in excess of $1/\bar{\tau}_H^c$, the
superconducting order parameter $\Delta_0(\vect{r})$ exhibits spatial
fluctuations around a zero mean with a correlation length
$L_D=\xi_0(\bar{\tau}_H^c/ \tau_H-1)^{-1/2}$ diverging at the critical
point. Regarding these spatial configurations as domains or
``droplets'' of ordered phase with a characteristic length scale
$L_D$, fluctuations $\delta\hat{\Pi}_0$ impose a long-ranged, complex,
random Josephson coupling --- a superconducting ``gauge
glass''~\cite{zhou_spivak}. While such a variational approach is
naively applicable close to the bulk transition --- see below --- one
might expect it to underestimate the impact of optimal fluctuations of
the random susceptibility.

Alternatively, treating the optimal fluctuations more accurately, one
can pursue a more direct `mean-field' approach, seeking explicit
solutions of the stochastic saddle-point equation~(\ref{eq:bcsco})
(cf.~Ref.~\cite{galitski_larkin}). In this case, guided by the
intuition afforded by the properties of band tail states of random
Hamiltonian operators, one might expect the behavior close to the
critical point to be dominated by spatially localized field
configurations of $\Delta_0(\vect{r})$ associated with rare or
\emph{optimal fluctuations} of the random operator $\hat{\Pi}_0$.
Ignoring the potential impact of spatial and dynamical fluctuations of
the order parameter around these saddle-point configurations, such an
approach would suggest that the formation of the homogeneous
superconducting condensate phase is preempted by the nucleation of
domains of ordered phase, the implication being that the bulk
transition will be mediated by phase ordering of Josephson coupled
islands. Crucially, since the droplets reflect fluctuations of the
complex pairing susceptibility, the condensate wave function would
itself be complex --- a superconducting `glass-like' phase of the type
described by Ref.~\cite{zhou_spivak}.

The distinction between the two approaches may to some extent be
semantic. While the former approach may underestimate the capacity of
the system to form `droplets' of condensate, the nature of the bulk
transition may rely more sensitively on the properties of the gauge
glass: the geometry and rigidity of the superconducting domains is
less important than the random Josephson coupling between the domains
arising from the fluctuations. However, in the parallel field
geometry, when taking into account the long-ranged nature of the
correlations of $\hat{\Pi}_0$, we will argue that the transition
affords a potentially different interpretation.  Specifically, we will
show that fluctuations $\delta\Pi_0({\bf r},{\bf r}')$ engage
\emph{long-ranged} diffusion modes and, as such, are correlated over
long distances. Such long-range correlations discriminate the spectral
properties of the pairing susceptibility from those of usual
short-ranged random impurity models. In particular, in contrast to
band tail states in low-dimensional Hamiltonian systems, the
transition between the tail state region and the `extended' bulk state
region is not mediated by a region of weakly localised states but
rather it is abrupt: Even states close to the band edge of
$\hat{\Pi}_0$ are power-law extended. If the condensate acquires the
texture of these states, they will in turn inherit `critical' spatial
correlations.

\begin{figure}
\begin{center}
  \includegraphics[width=1\linewidth,angle=0]{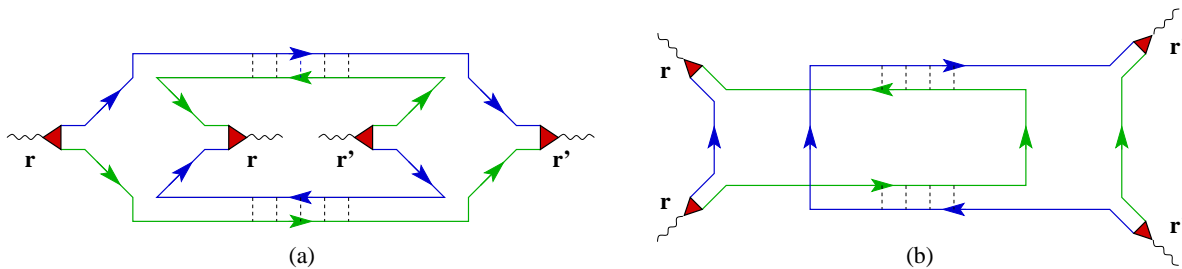}
\end{center}
\caption{\small Impurity diagrams providing the leading contribution
        to the variance of the matrix element
        $\delta\Pi_0(\vect{r},\vect{r}')$.  In both cases, the
        long-ranged character of the correlations can be ascribed to
        `diffuson' contributions which are insensitive to magnetic
        field. The influence of the magnetic field is imposed through
        the short-ranged `Cooperon' ladders which dress the vertices.}
\label{fig:corr_diag}
\end{figure}

To explore in detail the spectral properties of the (static) pairing
susceptibility $\hat{\Pi}_0$, one requires information about the full
distribution function of matrix elements, $P[\Pi_0]$. Clearly, being
of two-particle character, such a programme seems unfeasible --- at
least in the two-dimensional geometry considered here. Instead, we
will choose to characterize the scale of fluctuations through the
variance of $\Pi_0$ which can be estimated diagrammatically. Here, to
be consistent, one must discard fluctuations $\delta \Pi_0(\vect{r}
,\vect{r}')$ occurring on a length scale $|\vect{r}-\vect{r}'|$
shorter than the magnetic diffusion length $L_H=\sqrt{D\tau_H}$. Such
fluctuations, which effect equally matrix elements of the unperturbed
superconducting system, demand a more careful consideration of the
influence of disorder on the pairing interaction itself. These effects
are irrelevant to, and reach well-beyond, the present discussion.
Although the ensemble average has all but vanished, on length scales
$|\vect{r} - \vect{r}'| \gg L_H$, one finds that the matrix elements
of the pairing susceptibility become complex and exhibit significant
\emph{long-ranged fluctuations}. Taking into account the leading
contribution from the impurity diagrams depicted in
Fig.~\ref{fig:corr_diag}a, a diagrammatic estimate of their magnitude
obtains $\langle \delta\Pi_0 (\vect{r}_1 , \vect{r}_2) \delta \Pi_0
(\vect{r}_3 , \vect{r}_4)\rangle_V \simeq
\tilde\delta(\vect{r}_1-\vect{r}_4)
\tilde\delta(\vect{r}_2-\vect{r}_3)C(\vect{r}_1-\vect{r}_2)$, where
\begin{equation}
  C (\vect{r} - \vect{r}') \sim \Frac{\nu^2}{g^2}
  \left(\frac{\xi_0}{|\vect{r} - \vect{r}'|}\right)^4 \; ,
\label{eq:corre}
\end{equation}
with $g=\nu D$ the dimensionless conductance of the normal
two-dimensional system. Indeed, a similar result was obtained by
Spivak and Zhou~\cite{spivak_zhou} for the perpendicular field
geometry: the difference between the parallel and perpendicular field
orientations is manifest only in the short-ranged Cooperon
contributions which dress the vertices, giving rise to the envelope
functions of width $\sim L_H$ that we denoted by
$\tilde\delta(\vect{r})$ --- the long-range scaling is inherited from
the field-insensitive diffuson content.
While, for coordinates $\vect{r}_i$ separated by more than $L_H$,
matrix elements $\delta\Pi_0(\vect{r}_1 , \vect{r}_2)$ remain
statistically uncorrelated from those of $\delta\Pi_0(\vect{r}_3 ,
\vect{r}_4)$, it may be confirmed that the correlation function
$\langle \delta\Pi_0(\vect{r} , \vect{r}) \delta\Pi_0(\vect{r}'
,\vect{r}') \rangle_V$ (Fig.~\ref{fig:corr_diag}b) also scales as
$C(\vect{r} - \vect{r}')$.

Now, in the general two-dimensional system, the eigenfunctions of a
typical random Hermitian operator exhibit a continuous evolution from
strong localisation below the band edge of the non-disordered system,
to weak localisation above. Applied to the pairing susceptibility,
such behaviour would motivate a theory of the transition based on
phase ordering of domains large enough to survive the effects of
quantum fluctuations. However, these considerations overlook the
potential significance of the long-ranged power-law correlations of
$\delta\Pi_0(\vect{r},\vect{r}')$. In the present case, the complex
random matrix elements have a strength which decays only as a
power-law, with an exponent $\beta$ that coincides with the
dimensionality $D$, i.e. $\beta=D=2$ in the thin-film geometry.
Previous investigations by Levitov~\cite{levitov,altshuler_levitov}
have revealed that such a dependence places the system in a `critical
regime' where the bulk eigenstates of the susceptibility are neither
fully extended nor exponentially localized. Rather, such states
$\chi_{\alpha} (\vect{r})$ exhibit a multi-fractal structure all the
way down to the `band edge' with moments $\langle |
\chi_{\alpha}(\vect{r})|^{2q}\rangle_\Pi \sim L^{-2q + \eta_q}$
characterized by a set of exponents $\eta_q$, and power-law spatial
correlations,
\begin{equation}
  f (\vect{r} - \vect{r}') \equiv L^{4} \langle |\chi_{\alpha}^2
  (\vect{r}) \chi_{\alpha}^2 (\vect{r}') |\rangle_\Pi \sim
  \left(\Frac{L}{|\vect{r} - \vect{r}'|}\right)^{\eta_2} \; .
\end{equation}
At the level of the saddle-point, the transition to the \emph{bulk}
superconducting phase may not, after all, be a problem of optimal
fluctuations.

To make the analysis quantitative, one can use the
variance~(\ref{eq:corre}) to characterize the distribution of
fluctuations. On scales $|\vect{r} - \vect{r}'|\gg L_H$, let us
suppose that matrix elements are specified by a Gaussian distribution,
$P[\Pi_0] = \exp [-\frac12\int d\vect{r}\, d\vect{r}'\; | \delta
\Pi_0(\vect{r}, \vect{r}')|^2/C (\vect{r} - \vect{r}')]$, where we
include only the first type of correlations discussed above (see
Fig.~\ref{fig:corr_diag}(a)). On length scales smaller than $L_H$, we
will assume to be cautious that the matrix elements simply coincide
with their impurity average.  (Indeed, such an assumption
underestimates the renormalisation of the quantum critical point.)
Although there is no reason to expect the distribution to be Gaussian,
an estimate of the leading contribution to the higher cumulants is
compatible with the \emph{Ansatz} for $P[\Pi_0]$. With this
distribution, the rearrangement of the DoS implied by the long-ranged
fluctuations of $\hat{\Pi}_0$ can be estimated within the
self-consistent Born approximation (SCBA).  Defining the impurity
averaged Green function, $\hat{\mathcal{G}}_0=\langle (\gamma_+ -
\hat{\Pi} / \nu)^{-1}\rangle_\Pi$, where $\gamma_+\equiv\gamma+i0$,
the SCBA translates to the condition
\begin{equation}
  - \frac{g_0}{2\pi} \equiv \mathcal{G}_0 (\vect{r} , \vect{r})\simeq
  \int_0^{1/L_H} \Frac{d \vect{p}}{(2 \pi)^2}
  \frac{1}{\gamma_+-\bar\gamma_c + L_H^2 (\vect{p}^2/2+\alpha g_0)} \;
  ,
\label{eq:scbap}
\end{equation}
where (taking $\omega_D \tau_H \gg 1$) $\bar{\gamma}_c \equiv
\ln(\omega_D \tau_H)$, and $\alpha = \kappa/ g^2 (\Delta_0 \tau_H)^2$
characterizes the strength of fluctuations, where $\kappa$ is some
numerical constant. Setting $\alpha=0$, the self-consistent equation
recovers the constant DoS of the ensemble averaged operator. In this
approximation, as expected from the AG condition, the transition to
the superconducting phase takes place into a state with a spatially
homogeneous order parameter when $(\lambda_{\rm BCS}\,\nu)^{-1} -
\bar{\gamma}_c \equiv \ln(2/\Delta_0\bar{\tau}_H^c) =0$. Reinstating
mesoscopic fluctuations (through $\alpha$), one finds the renormalized
edge $\gamma_c\simeq \bar{\gamma}_c+\alpha$, valid in the limit
$(\gamma_c-\bar{\gamma}_c)/\bar{\gamma}_c \ll 1$, from which one can
infer the shift of the critical field,
\begin{equation}
  \frac{1}{\tau_H^c} \simeq \frac{1}{\bar{\tau}_H^c}\left(1 +
  \Frac{\kappa}{4 g^2}\right)\; .
\end{equation}
An expansion of the DoS in the vicinity of $\gamma_c$ obtains
$\rho(\gamma) = \langle \tr \delta (\gamma -\hat{\Pi} /\nu)\rangle_\Pi
\simeq (3/\pi^2\alpha)^{1/2}\sqrt{\gamma_c-\gamma}$. Note that
inclusion of the $\langle \delta\Pi_0(\vect{r} , \vect{r})
\delta\Pi_0(\vect{r}' ,\vect{r}') \rangle_V$ correlations leads only
to a change in $\kappa$.

Therefore, leaving aside the potential for a small exponential band of
Lifshitz tail states of $\hat{\Pi}_0$, at the level of the linearised
saddle-point approximation, the onset of superconductivity at $T=0$
takes places at a value of $1/\tau_H^c$ renormalized from that
predicted by AG theory. The bulk transition is to a glass-like phase
in which the \emph{complex} order parameter inherits the multi-fractal
structure implied by the critical theory,
\begin{equation}
  \langle |\Delta (\vect{r},t)|^2
  |\Delta(\vect{r}',t')|^2\rangle_{\Pi,\Delta}- \langle
  |\Delta(\vect{r},t)|^2\rangle_{\Pi,\Delta}^2 \sim
  f(\vect{r}-\vect{r}') \; .
\end{equation}

The integrity of the saddle-point approximation depends sensitively on
the impact of dynamical and spatial fluctuations of the order
parameter.  Their significance can be inferred from the behaviour of
the impurity averaged susceptibility. The gradient expansion, $\langle
\Pi_{\omega_m}(\vect{r},\vect{r}')\rangle_V\simeq \nu\delta(\vect{r}
-\vect{r}')[\gamma_c+L_H^2\partial_{\vect{r}'}^2/2-|\omega_m|\tau_H/4\pi]$,
shows the system to be dissipative (both outside and within the
gapless ordered phase) with a rate proportional to $1/\tau_H$ implying
a dynamical exponent $z=2$. In the two-dimensional geometry, this
places the system at its upper critical dimension~\cite{Hertz}. In the
presence of mesoscopic fluctuations, the phase space for low-energy
fluctuations is only diminished justifying the saddle-point analysis
adopted in this work.

At temperatures $T \ne 0$, the thermodynamic properties of the system
demand two further considerations: As well as the obvious significance
of thermal fluctuations of the order parameter, the mechanisms of
quantum interference, on which the long-range correlations of the
pairing susceptibility rely, become gradually extinguished. Thermal
smearing limits the long-range correlations of
$\delta\Pi_0(\vect{r},\vect{r}')$ to length scales $|\vect{r}-
\vect{r}'|$ smaller than the thermal length $L_T=\sqrt{D/T}$, i.e.
$C(\vect{r} - \vect{r}') \sim (\nu^2/g^2)(\xi_0/|\vect{r} -
\vect{r}'|)^4 e^{- |\vect{r} -\vect{r}'|/L_T}$. Providing $L_T$
remains greatly in excess of $L_H$, such dependence motivates a
coarse-grained description: Over the range $1/\bar{\tau}_H^c <
1/\tau_H <1/\tau_H^c$, each domain of size $L_T$ has condensed into a
glass-like superconducting droplet, weakly connected to neighboring
domains by the residual fluctuations of $\delta\hat{\Pi}_0$. On length
scales in excess of $L_T$, one therefore expects to recover a ``gauge
glass'' picture analogous to that envisaged by Spivak and
Zhou~\cite{zhou_spivak} with a low-temperature behaviour that can be
inferred from the analysis of Galitski and
Larkin~\cite{galitski_larkin}.

\emph{Acknowledgments}: We are indebted to Alex Kamenev and Boris
Spivak for useful discussions. FMM would like to acknowledge the
financial support of EPSRC (GR/R95951). JSM was supported by a Feodor
Lynen fellowship of the Humboldt Foundation.

\end{document}